\documentstyle[aaspp4]{article}

\def\k{{\bf k}}
\def\r{{\bf r}}

\def\u{{\bf u}}
\def\uzk{u_{\bf k}}
\def\x{{\bf x}}
\def\z{{\bf z}}

\begin{document}
\title{Emissivity Statistics in Turbulent, Compressible MHD Flows and
the Density-Velocity Correlation}
\author {Alex Lazarian$^1$, Dmitri Pogosyan$^2$, Enrique V\'azquez-Semadeni$^3$, and B\'arbara Pichardo$^4$}

\affil{$^1$Dept.\ of Astronomy, University of Wisconsin, Madison, USA}
\affil{$^2$Canadian Institute for Theoretical Astrophysics, University
of Toronto, CANADA}
\affil{$^3$Instituto de Astronom\'\i a, UNAM, Campus Morelia,
Apdo. Postal 3-72, Xangari, 58089, Morelia, Mich., MEXICO} 
\affil{$^4$Instituto de Astronom\'\i a, UNAM, Apdo.\ Postal 70-264,
M\'exico D.F.\ 04510, MEXICO}

\begin{abstract}

In this paper we test the results of a recent analytical study by
Lazarian and Pogosyan, on the statistics of emissivity in velocity
channel maps, in the case of realistic density and velocity fields
obtained from numerical simulations of MHD turbulence in the
interstellar medium (ISM).
To compensate for the lack of
well-developed inertial ranges in the simulations due to the limited
resolution, we apply a procedure for modifying the spectral slopes of
the fields while preserving the spatial structures.
We find that the
density and velocity are moderately correlated in space
and prove
that the analytical results by Lazarian and Pogosyan
hold in the case when these fields obey the fluid conservation
equations. Our results imply that the spectra of velocity and density can
be safely recovered from the position-position-velocity (PPV) data cubes
available through observations, and confirm that the relative
contributions of the velocity and density fluctuations to those of the
emissivity depend on the velocity resolution used and on the steepness
of the density spectral index. Furthermore, this paper supports
previous reports that an
interpretation of the features in the PPV data cubes as simple density
enhancements (i.e., ``clouds'') can be often erroneous, as we observe that
changes in the velocity statistics substantially modify the statistics
of emissivity within the velocity data cubes.

\end{abstract}


\section{Introduction}

It is generally accepted that interstellar turbulence plays a crucial
role in many astrophysical processes, including molecular cloud and
star formation, mass
and energy transfer in accretion disks, acceleration of cosmic rays,
etc. At present we
are still groping for the basic facts related to this complex
phenomenon and one of the reasons for such an unsatisfactory state of
affairs is that the interstellar turbulence statistics are not
directly available. For instance, in studies of the neutral medium,
indirect measures, such as spectral line widths and centroids of
velocity, are employed (e.g. Miesch \& Bally 1994; see also the review
by Scalo 1987), while potentially more valuable sources of information
available in velocity-channel maps remain mostly untapped (although
see Heyer \& Schloerb 1997; Rosolowsky et al.\ 1999; Brunt \& Heyer
2000) because of the difficulty of relating the
two-dimensional (2D) statistics available through observations to the
underlying three-dimensional (3D) statistics. A discussion of various
approaches to the problem of turbulence study using spectral line data
can be found in a recent review by Lazarian (1999).

In particular, the problem has been recently addressed by Lazarian
\& Pogosyan (2000, herafter LP00), who derived the index (i.e., the
logarithmic slope)
of the power spectrum\footnote{Note that throughout this paper we
refer exclusively to the {\it spatial} power spectrum of the various
fields, i.e., the Fourier transform of their auto-correlation
function, as is common in turbulence studies. Such spectra should not
be confused with, for example, spectral line profiles, or the spectra
of time series of data, which we do not consider here. Also,
throughout this paper, we stick to the convention that the spectral index of an
$N$-dimensional field does {\it not} include the $k^{N-1}$ factor
corresponding to a summation over a shell of wavenumbers of radius
$k$. With this notation, the well-known Kolmogorov $-5/3$ law corresponds
to an index of $-11/3$.} of the
intensity in velocity channel maps as a function of
the corresponding indices for the 3D density and velocity fields of
the emitting medium. Their work effectively provides a means of
inverting the problem, so that the power spectra of the medium's
density and velocity can be retrieved from the power spectrum of the
emissivity. Note that this procedure involves considering channels of
various velocity widths or ``velocity slices of different thickness'',
in the terminology of LP00.
Otherwise there is an indetermination due to the fact
that for sufficiently shallow density spectra the emissivity
in slices is given by a linear combination of the velocity and
density indices. Recent measurements by Stanimirivic (2000) and
Stanimirivic \& Lazarian (2000) have confirmed
a variation of the spectral index of HI intensity maps as the slice
width is varied, in accordance with the predictions of LP00.
The approach suggested by LP00 allows, in principle, to use the
wealth of existing spectroscopic data for deriving interstellar turbulence
statistics, thus possibly permitting new levels of understanding of
this phenomenon.
However, the derivation of LP00 assumes that the statistics of
velocity and density are independent. Although those authors showed that for
a particular case their results hold even when maximal correlation
of velocity and density is assumed, testing the scheme on realistic fields
from compressible magneto-hydrodynamic (MHD) simulations is
essential.

In this paper we assess the degree of correlation between the
density and velocity fields from MHD, compressible turbulence
simulations, and whether the theoretical results from LP00 apply in
this case, our goal being to find out to what extent the spectra of
velocity and density can be recovered from the observed emissivity
statistics. Note that
if the interdependence of velocity and density is important
for the recovery of their spectral indices, the inversion
must be modified to account for the velocity-density
correlations.

Numerous measurements of the emissivity within actual data cubes
suggest that the spectrum follows a power law (Green 1993, Stanimirovic et
al.\ 1999), as is the case in classical high-Reynolds number
incompressible turbulence (e.g., Lesieur 1990), and high-resolution
numerical simulations of highly compressible MHD turbulence in less
than three dimensions (e.g., Passot, V\'azquez-Semadeni \& Pouquet
1995; Gammie \& Ostriker 1996). A power-law assumption
was also used by LP00. Numerical simulations of the ISM in 3D, however, usually
do not have a large enough inertial range to produce good power laws
and this can complicate our interpretation of the results. To
deal with this issue, below we describe a procedure for ``correcting''
the spectral indices of the numerically-simulated density and velocity
fields while preserving their underlying spatial correlations.

We describe the way how our numerical data were produced, and the
correlation between density and line-of-sight velocity in section
\ref{sec:num_exp}, compare numerical results with the predictions of
LP00 in section \ref{sec:stats}, and provide a general discussion in
section \ref{sec:disc}. Finally, our
conclusions are summarized in section \ref{sec:sum}.

\section{Numerical Data} \label{sec:num_exp}

\subsection{Simulations and Spectrum Modification} \label{sec:sims}

In order to test whether the results from the analytical study of
LP00 carry over to the case
when the density and velocity fields are correlated according to
self-consistent fluid evolution, in the following sections we explicitly
calculate the spectral slopes for velocity channel-map data in
position-position-velocity (PPV) cubes obtained
from a three-dimensional simulation of the ISM at intermediate size
scales (3--300 pc). The simulation is nearly identical to the one presented by
Pichardo et al.\ (2000), except at slightly lower resolution. Both the
simulation and the procedure for obtaining the velocity channel data
have been discussed at length in that paper, and we refer the reader
to it for details. In this section we just outline the information
necessary for the purposes of the present paper. Channel
maps are essentially maps of column density within a given
line-of-sight (LOS) velocity interval. Throughout this paper we refer
to such column density as ``emissivity'' or ``map intensity'', in
analogy to the
observational situation. For the sake of simplicity,
thermal broadening, whose effects on the statistical analysis are
discussed by LP00, is not considered.

The simulation represents a cubic box of 300 pc on a side on the
Galactic plane, centered at the solar Galactocentric distance. The
magneto-hydrodynamic equations are solved on a $100^3$ Cartesian grid
with the $x$, $y$ and $z$ directions respectively representing the
radial, tangential and vertical directions in the Galactic disk. A
pseudospectral scheme with periodic boundary conditions is used, which
includes self-gravity, parameterized
heating and cooling, and modeled star formation, such that a ``star''
(a local heating source which causes its surroundig gas to expand) is
turned on wherever the density exceeds a
threshold $\rho_{\rm c}$ and $\nabla \cdot {\bf u} < 0$. The ``stars''
remain on for 3.7 Myr after the criterion is met. This procedure is
intended to mimic HII region expansion from OB star ionization
heating, and provides an energy source for maintaining the turbulence
in the simulation. Energy injection by supernovae is not included due
to limitations of the numerical scheme, but the total energy injected
per stellar event over the stellar lifetime is of the same order of
magnitude as that which would be injected by a supernova event. The
simulations also include the Coriolis force corresponding to a
rotation of the frame around the Galactic center every $2 \times 10^8$
yr, and a shear in the $x$-$y$ plane of the form $u_0 = 2.4 
~\sin (2 \pi y/300~{\rm pc})$ km s$^{-1}$, where $u$ is the
$x$-component of 
the velocity field. The turbulent motions occur on top of this
shearing velocity. This sinusoidal shear is not
highly realistic, but is the only
one possible with periodic boundary conditions, and was introduced
by Passot et al.\ (1995) as a
crude approximation of the galactic shear. Due to the periodic
boundary conditions, no stratification is present in the
$z$-direction, but this is not too strong a concern given the range of
scales represented by the simulation.

\begin{figure}
\plottwo{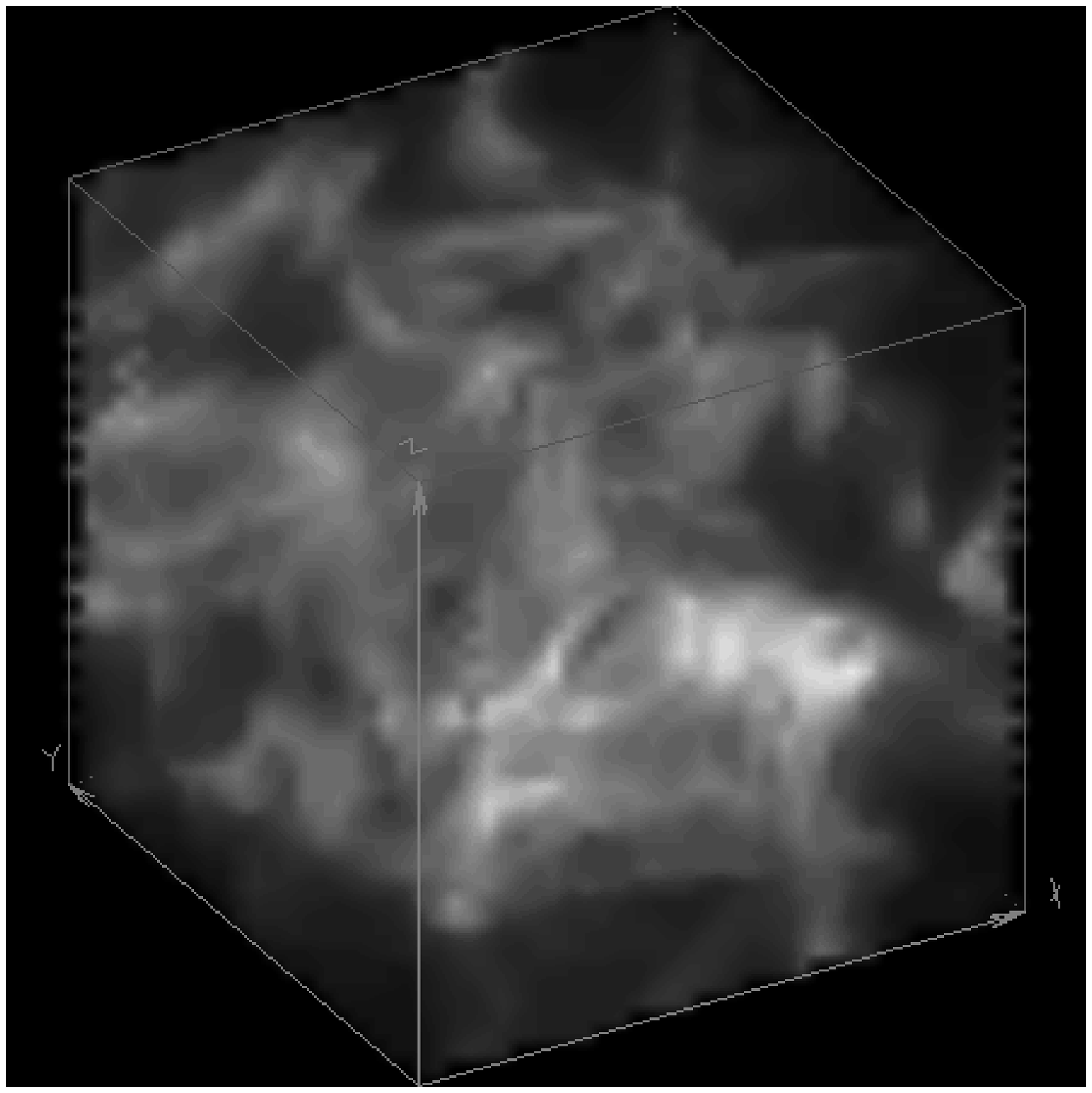}{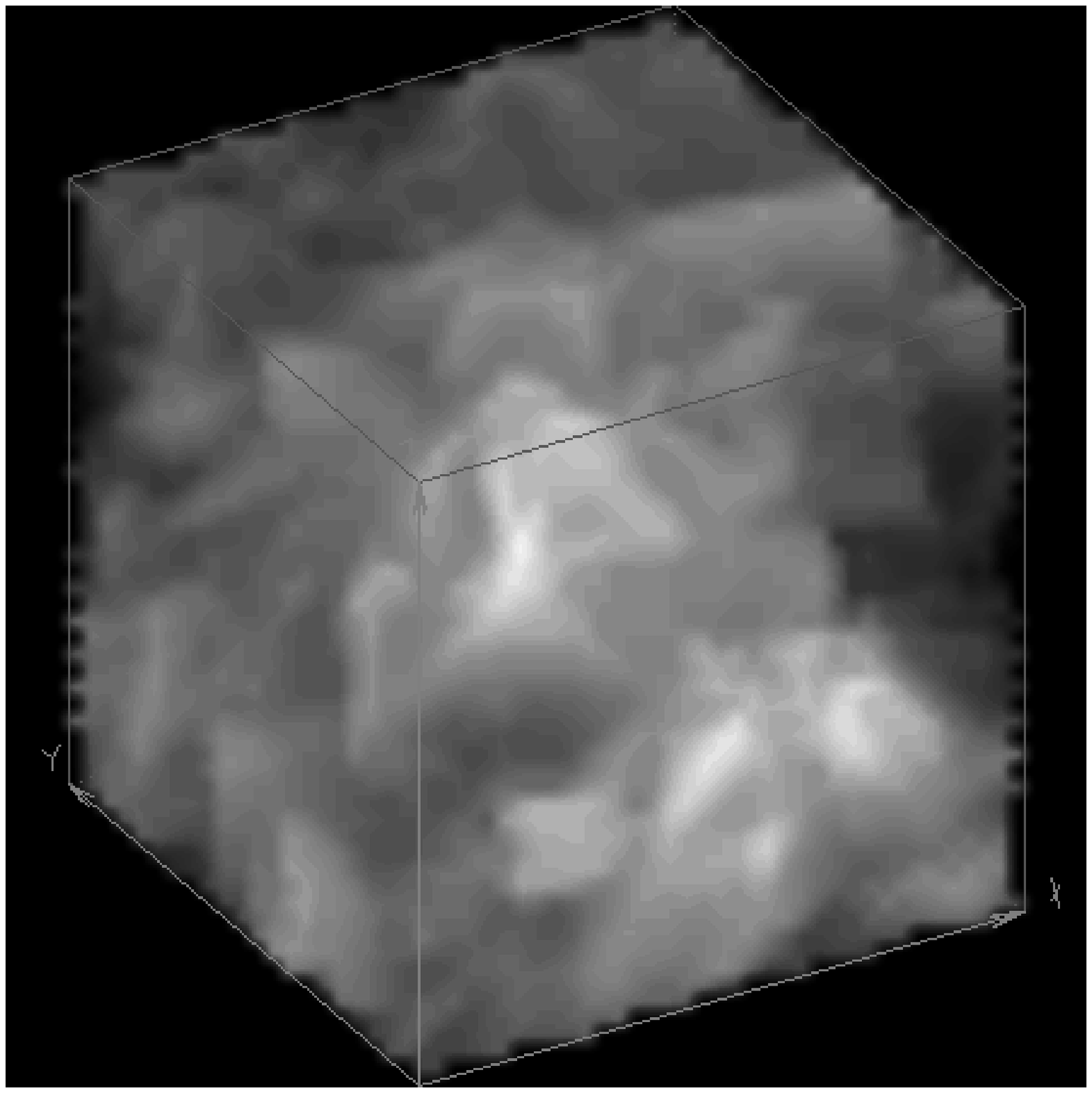}
\caption{{\it a) (Left Panel)} Three-dimensional (3D) logarithmic image of the
original density field. The maximum and minimum density values are
$109$ and 0.43 cm$^{-3}$. {\it b) (Right panel)} 3D image of the
LOS-velocity field, i.e., the $z$-component of the velocity vector. In
both cases, whiter color means larger values.} 
\label{fig:den_uz}
\end{figure}

Because the star formation prescription prevents the density from
reaching values significantly larger than $\rho_{\rm c}$, we turn it
off after the turbulence is fully developed, and focus on a snapshot
of the simulation shortly after that time. At the time of the
snapshot, the maximum and minimum values of the (number) density are 109 and
0.43 cm$^{-3}$, respectively. A uniform magnetic field of 1.5 $\mu$G
parallel to the $x$-direction is included, on top of which turbulent
magnetic fluctuations induced by the stellar energy injection
occur. The maximum and minimum values of the magnetic field strength
are 12.5 and $4.5 \times 10^{-2} \mu$G. An image of the density field
is shown in fig.\ \ref{fig:den_uz}a.  We adopt the $z$-direction in
the simulation (perpendicular to the Galactic plane) as the LOS
direction, in order to prevent the shear from introducing power
unrelated to the turbulent fluctuations into the velocity
spectrum. Other than that, since the simulation has no
vertical stratification, the $z$ direction is statistically equivalent
to the $y$ direction. Only the $x$-direction, parallel to the mean
magnetic field, is special. Figure
\ref{fig:den_uz}b shows an image of the LOS-component of the velocity
field. The structures in
the simulations have been discussed at length by Pichardo et al.\
(2000).

However, neither the density nor the LOS-velocity fields have spectra
well suited for studying the emissivity spectrum since, due
to the low
resolution of the simulation, the spectra are not clear power laws,
while most turbulence theories, including LP00's, consider power-law spectra.
To circumvent this problem, in what follows we use modified density
and velocity fields, such that their spectra are indeed power laws,
but the phase coherence of the original fields is preserved.
A partial justification for this procedure stems from the fact
that the spatial information is contained in the phases of the
Fourier decomposition of a given field, while the spectrum is related
exclusively to the relative amplitudes of the various modes (Armi \&
Flament 1985). Unfortunately, this justification is not complete, because
the velocity-density coherence does change to some extent as the spectral
slope is modified. We discuss this issue in some more detail
in \S \ref{sec:correl}.

As an indication of what would be realistic indices for the fields, we
note that incompressible MHD simulations
that resolve the inertial range (Cho \& Vishniac 2000a,b) tend to
result in a Kolmogorov-type spectrum with slope $-11/3$,
as theoretically predicted
by Goldreich \& Shridhar (1995), while highly compressible simulations
in less than 3D (e.g., Passot et al.\ 1995; Scalo et al.\ 1998 [2D];
Gammie \& Ostriker 1996 [1+2/3 D])
tend to give shock-spectra for the velocity with slopes near $-4$ and
density spectra with slopes near $-2$. Thus, one may expect
spectral indices close to these values
also in 3D highly compressible turbulence, so that
the spatial correlations from our simulations should probably be most
appropriate for those ranges of values.
On the other hand, we do not know beforehand
what sort of phase coherence should be present for either
shallower or steeper spectra, and therefore in these cases our present
study is limited to testing whether the formulae by LP00 are correct
in the presence of the particular sort of correlations produced by our
simulations.

\begin{figure}
\plotone{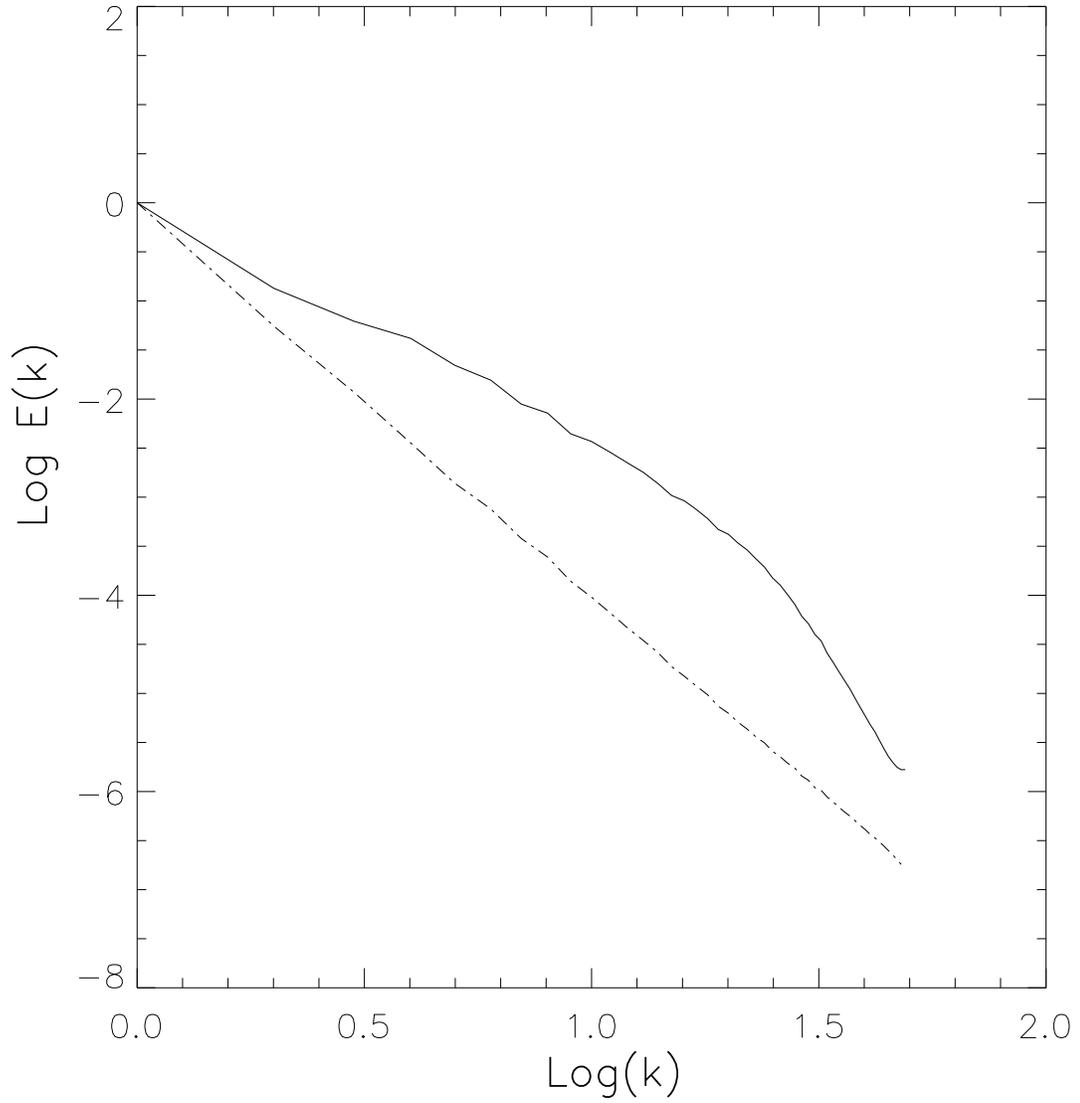}
\caption{Spectra of the density field in original form ({\it solid line}) and
after modification to an index of $-4$ ({\it dash-dotted line}). The
wavenumber $k$ here is defined as $L/\lambda$, where $\lambda$ is the
wavelength associated with $k$, and $L$ is the box size.}
\label{fig:old_new_spec}
\end{figure}

\begin{figure}
\plottwo{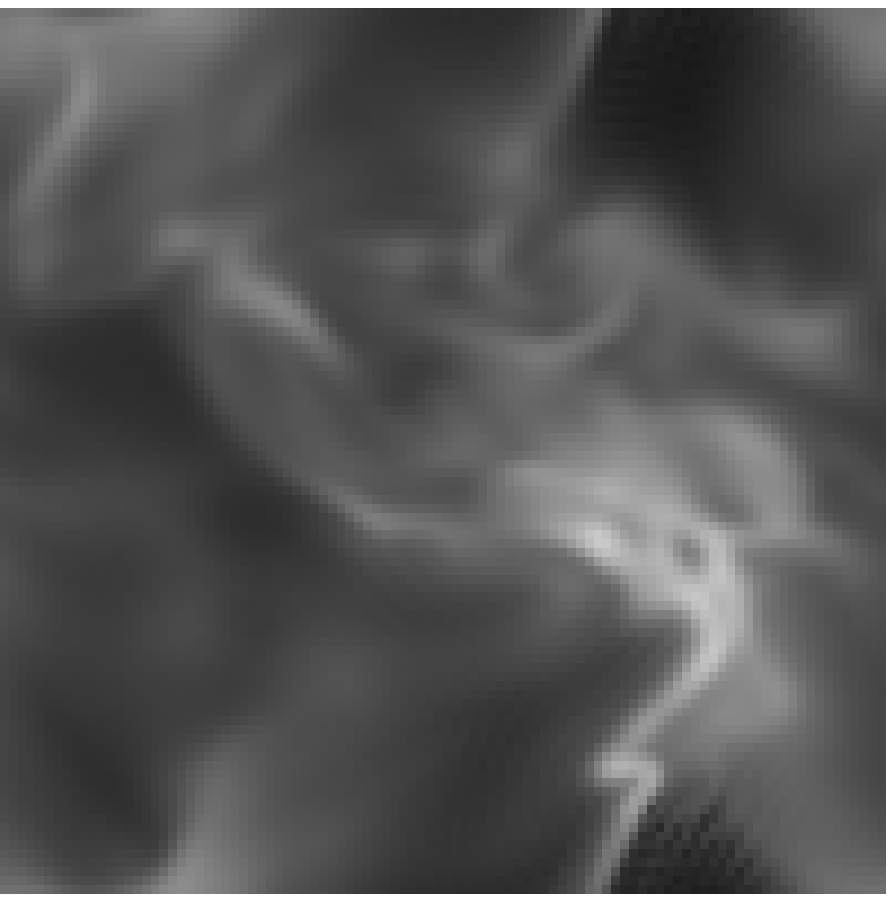}{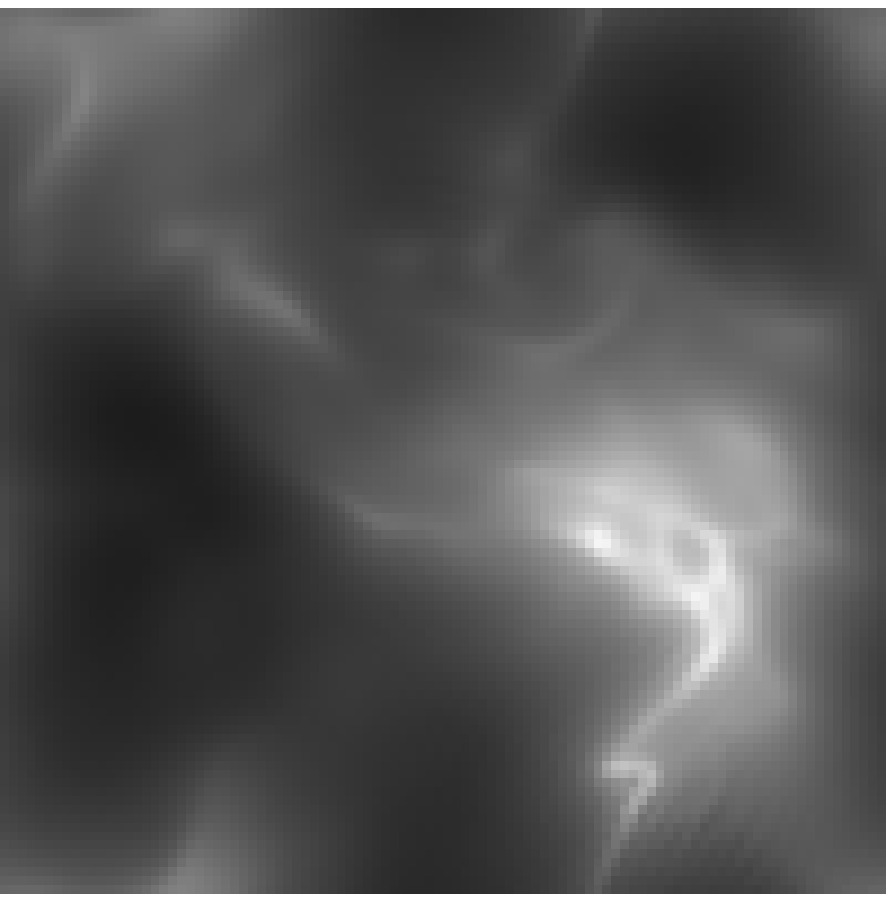}
\caption{Two-dimensional cuts through the 3D density field in original
form ({\it left panel}) and after modification to a spectral index of
$-5$. Note that the modified-spectrum image appears smoother, because the
relative importance of the small scales has been reduced in this case.}
\label{fig:old_new_dens}
\end{figure}

The spectral index modification procedure is as follows. We perform a
Fourier decomposition of the density and LOS velocity as
\begin{eqnarray}
&&
\rho(\x) = \sum_\k \rho_\k  \exp \big(i \k \cdot \x \big) \\
&&
u_z(\x) = \sum_\k \uzk \exp \big(i \k \cdot \x \big),
\label{eq:Fourier_dec}
\end{eqnarray}
where \x\ is the position vector, $\rho_{\bf k}$ and $u_{\bf k}$ are
the Fourier
amplitudes of the fields (we drop the subindex $z$ in the latter for
convenience) and \k\ is the (vector) wavenumber. In general, the
amplitudes are of the form $X_\k = |X_\k| e^{i \phi}$,
where $|X_\k|$ and $\phi$ are respectively the magnitude and phase of
$X_\k$. Thus, we replace the magnitudes of both fields by new ones
(labeled by primes) satisfying $|{X_\k}^\prime|^2 \propto k^n$, where
$n$ is the desired power-law index. The phases are
unchanged. The modified amplitudes are then transformed back to
configuration space, rendering new fields which preserve the spatial
structure of the original fields but with the desired power-law
spectrum. We avoid aliasing by including only modes with wavenumbers
lying within a circle of radius equal to half the resolution in
wavenumber space in the inverse transform.
As an example, in fig.\ \ref{fig:old_new_spec} we show the
original power spectrum of the density field together with the
resulting modified spectrum for $n=-4$, while fig.\ \ref{fig:old_new_dens}
shows cuts through the original and modified density fields, to
appreciate the kind of structural changes arising from the
spectrum-modification procedure. In the remainder of this paper, we reserve 
the symbol $n$ for the index of the density power spectrum, and denote 
the velocity power (or energy) spectrum index by $\mu$.

Using this procedure, we construct a suite of density fields with
spectral indices $-2.5$, $-4$, and $-5$, and of velocity fields with spectral
indices $-3.2$, $-3.7$ (the Kolmogorov value), and $-4.5$. From these,
we then construct the corresponding PPV cubes for each
density-velocity pair, which are used in the next section to
obtain the projected intensity and velocity-channel maps.

\subsection{Density-Velocity correlation} \label{sec:correl}

To characterize the correlation between velocity and density one
can use the following function
\begin{equation}
C(\r)=\frac{\langle \rho(\x)\ \u(\x+\r)\cdot\r/|r|\rangle}{\sigma_\rho
\sigma_{\u}},
\label{eq:LP00_corr}
\end{equation}
where \r\ is the spatial separation (or ``lag''), \x\ is the spatial
position, $\rho$ is the density field, \u\ is the total
three-dimensional velocity vector, and
$\sigma_\rho$ and $\sigma_{\u}$ are respectively the
standard deviations of the density and of the velocity.
The averaging is performed over all \x-space and over all angles of
the lag vector \r.
This function is similar to the function $F({\bf r})$ introduced
in LP00 to calculate the effect of maximal allowable velocity-density
 correlations. The difference between eq.\
(\ref{eq:LP00_corr}) and  $F({\bf r})$ amounts
to the normalization and the $\cos\theta$ dependence, where $\theta$ is the
angle between ${\bf r}$ and the $z$-axis.

\begin{figure}
\plottwo{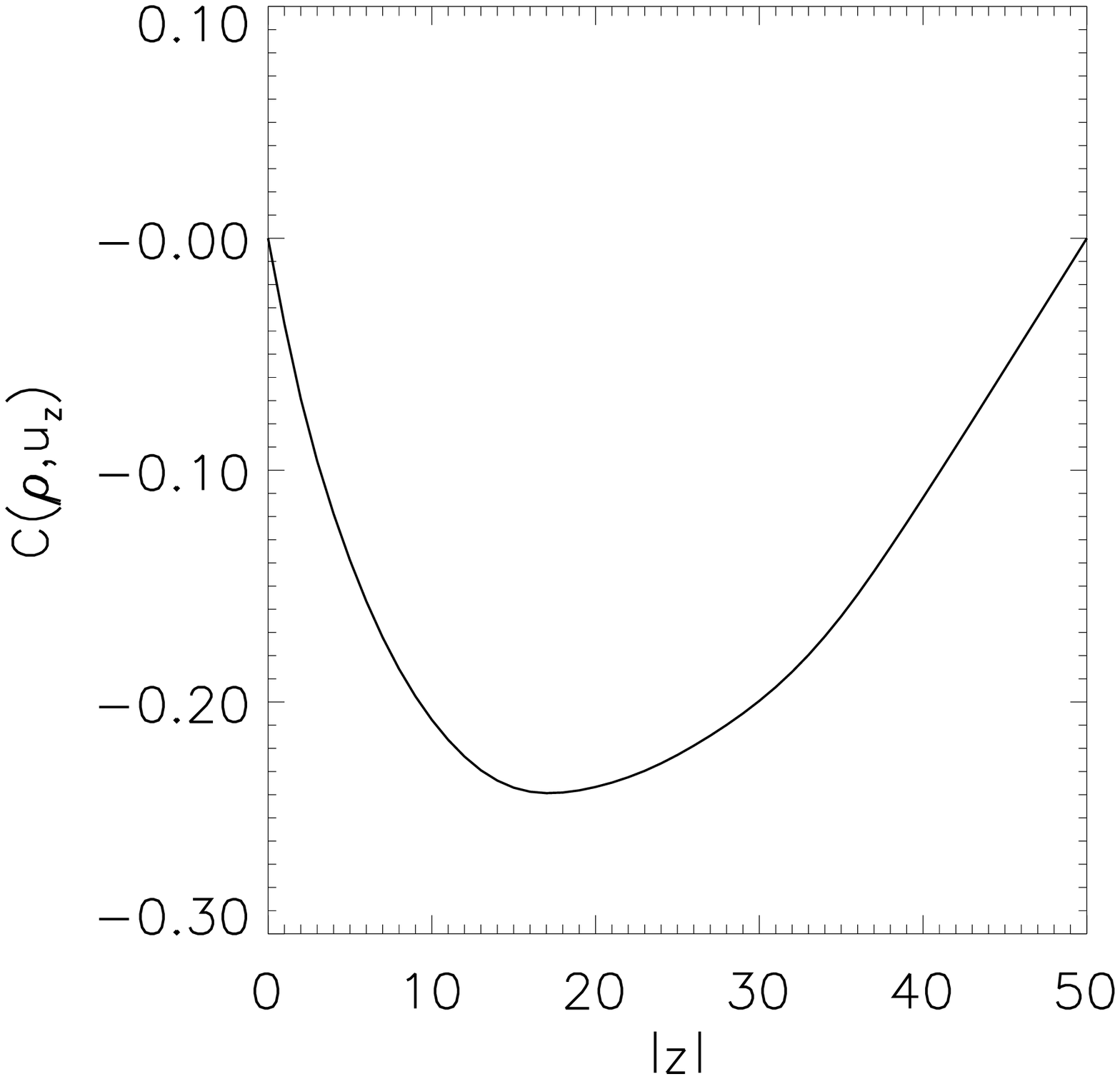}{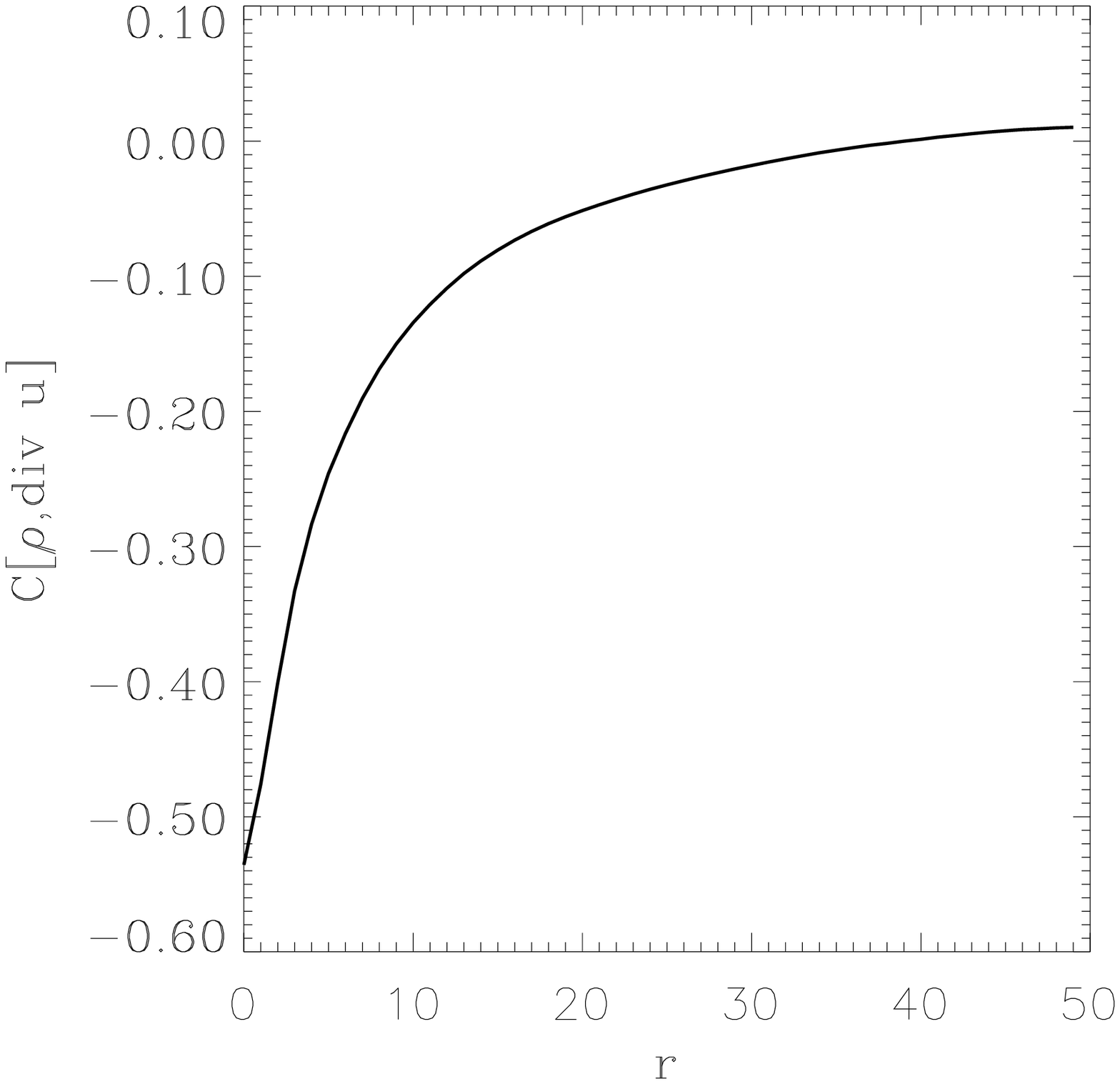}
\caption{Cross-correlation between the density and the line-of-sight
projection
velocity fields ({\it a), left panel}), and between the density and the
divergence of the total 3D velocity field ({\it b), right panel}). The
density is significantly correlated with the velocity divergence, while
the density-velocity correlation is weaker. The non-monotonous character
of the correlation defines a range of scales at which
the density-velocity correlation is maximal and the LP00 analysis
requires testing.}
\label{fig:correl}
\end{figure}

However, since the projection along \r\ precludes computing this
correlation using Fourier transform techniques, and computing the full
correlation numerically in
physical space would take prohibitely long times, in this paper we
compute a related
correlation involving only the LOS components of both the lags and of
the velocity vector. This
correlation, given by
\begin{equation}
C(|z|)=\frac{\langle \rho(\x)\ {\bf u}(\x+\z)\cdot\z/|z|\rangle}{\sigma_\rho
\sigma_{u_z}},
\label{eq:cross_corr}
\end{equation}
is much more economical numerically, but contains the same
essential information as eq.\ (\ref{eq:LP00_corr}). In
eq.\ (\ref{eq:cross_corr}), $\z=(0,0,z)$ is the separation along the
LOS and the
averaging is performed over all positions $\x$ in the simulation
box and over separations $z$ and $-z$. This correlation is shown in
fig.\ \ref{fig:correl}a for the original density and LOS-velocity fields.

As mentioned in \S \ref{sec:sims}, the correlations do change to some
extent upon the change in spectral indices. We have observed that the
correlation increases as the slopes of either the density or the
velocity spectra are made steeper. Since a steeper (resp.\ shallower)
slope means that the small scales are depressed (resp.\ enhanced) with
respect to the large scales, the observation implies that the density
and velocity fields are more similar to each other at large scales
than at small 
ones. The correlation between the modified fields most resembles that
between the original fields when the modified slopes are closest to
those of the original spectra.

The correlations observed are moderate,
having a maximum $\sim 0.25$ (in absolute value) for the original fields, and
$\sim 0.4$ for the modified fields with steepest slopes.
In retrospect, this result can be understood in terms of the fact that
the fluid equations link one field with the spatial derivatives of the
other: in the continuity equation, the density evolution is determined
by the {\it divergence} of the velocity field, and in the momentum
equation, the velocity evolution is linked to the pressure {\it
gradient}, which at best can be related to the density gradient, if
the flow behaves in an approximately barotropic way (e.g.,
V\'azquez-Semadeni, Passot \& Pouquet 1996). Thus, stronger correlations
may be expected between the density and the velocity divergence, and
between the velocity and the density gradient, but not so much between the
plain density and LOS-velocity fields. Indeed, fig.\ \ref{fig:correl}b
shows the correlation between the original density field and the divergence of
the original 3D velocity field. A  larger (negative) correlation, of
up to $-0.53$ is seen, supporting this interpretation.

Whether or not a density-velocity correlation of the observed strength
can alter the results of LP00 is not clear {\it a priori}, and this is the
motivation for our study below.

\section{Statistics of Velocity-Channel Maps} \label{sec:stats}

Unlike the observational situation, in numerical simulations
the density and velocity fields are available directly. This
allows us to simulate observations and to control the accuracy
with which the individual 3D statistics are recovered.

Spectroscopic observations usually deal with velocity channel maps
in which the velocity resolution is determined by the instrument.
In the case of our numerical simulation, velocity resolution
is not an issue, and we can produce as many velocity channels as desired
from the original density and LOS-velocity data cubes. Surely
the statistics within an individual slice
are degraded as the slice gets thinner, producing progressively higher
levels of shot noise, but we compensate for this
effect by averaging over the whole set of slices available.

To compare our results with the theoretical predictions of LP00 we recall
that two possible regimes were found there, depending on the width
of the velocity channels. They were termed
``thin'' and ``thick'' slicings, and their emergence is well motivated
physically. If we consider turbulence
at a scale $l$, the expected squared velocity difference (the
second-order structure function) between points
separated a distance $l$ scales as
$Cl^{m}$ , where $m$ is $2/3$ for
Kolmogorov-type turbulence,
and $C$ is a constant that depends on the intensity of
turbulence (see, e.g., Lesieur 1990). The structure function
index $m$ is related to the velocity power spectrum's index $\mu$ by
$m=-\mu-3$. Thus, when this velocity difference is larger
than the width
of the velocity slice, the slice is considered {\it thin}, and
{\it thick} otherwise.
LP00 showed that different slopes for the emissivity spectrum
are expected  in the two regimes mentioned above. It is obvious that
whether the slice is either thick or thin with respect to the
characteristic turbulent velocity difference depends on the scale
being considered, and
therefore, since the thin and thick asymptotics differ in general,
we expect to see a change of the slope of the emissivity spectrum at
some transition scale which depends on the channel width.

\begin{figure}[ht]
{\centering \leavevmode
\epsfxsize=.25\columnwidth \epsfbox{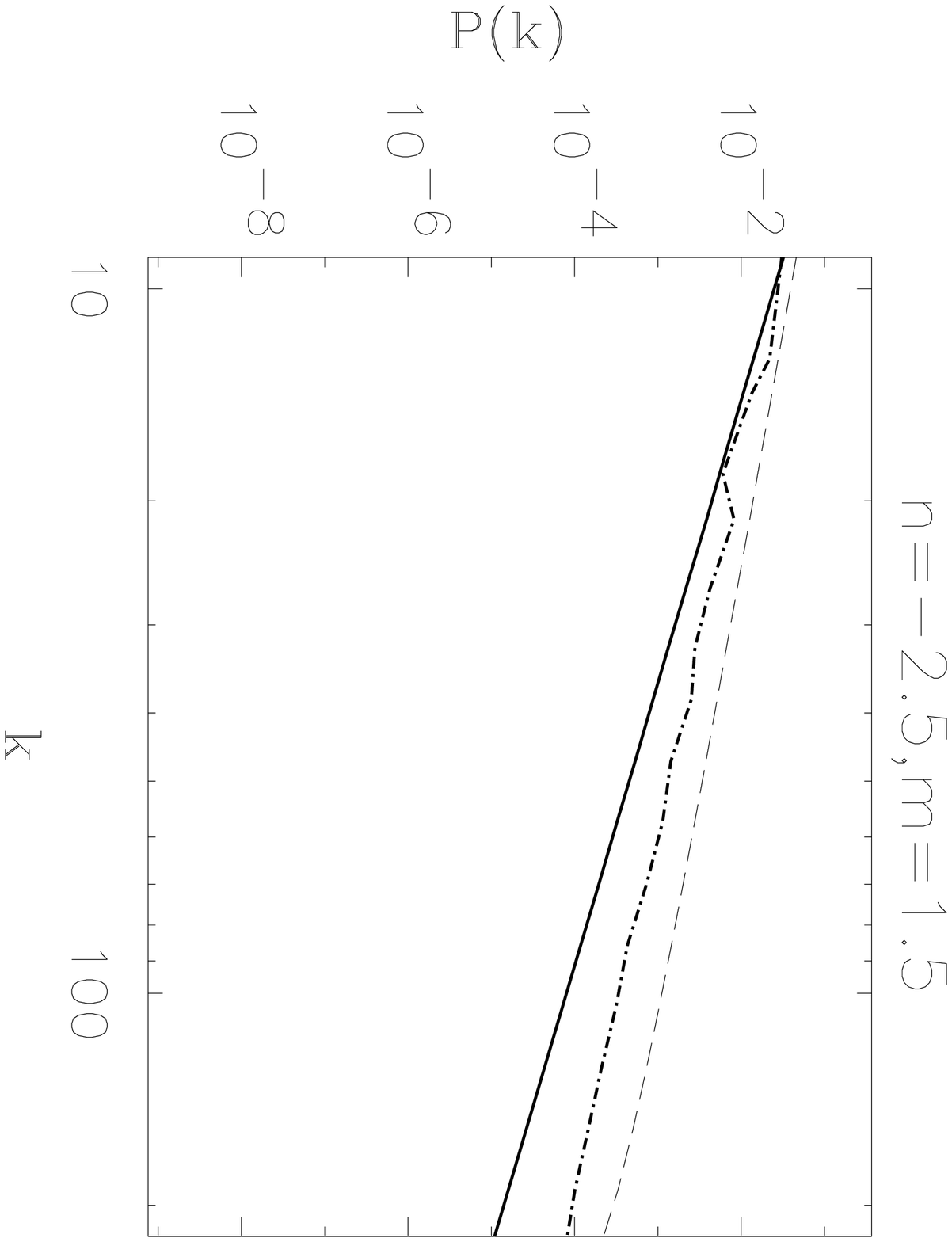} \hfil
\epsfxsize=.25\columnwidth \epsfbox{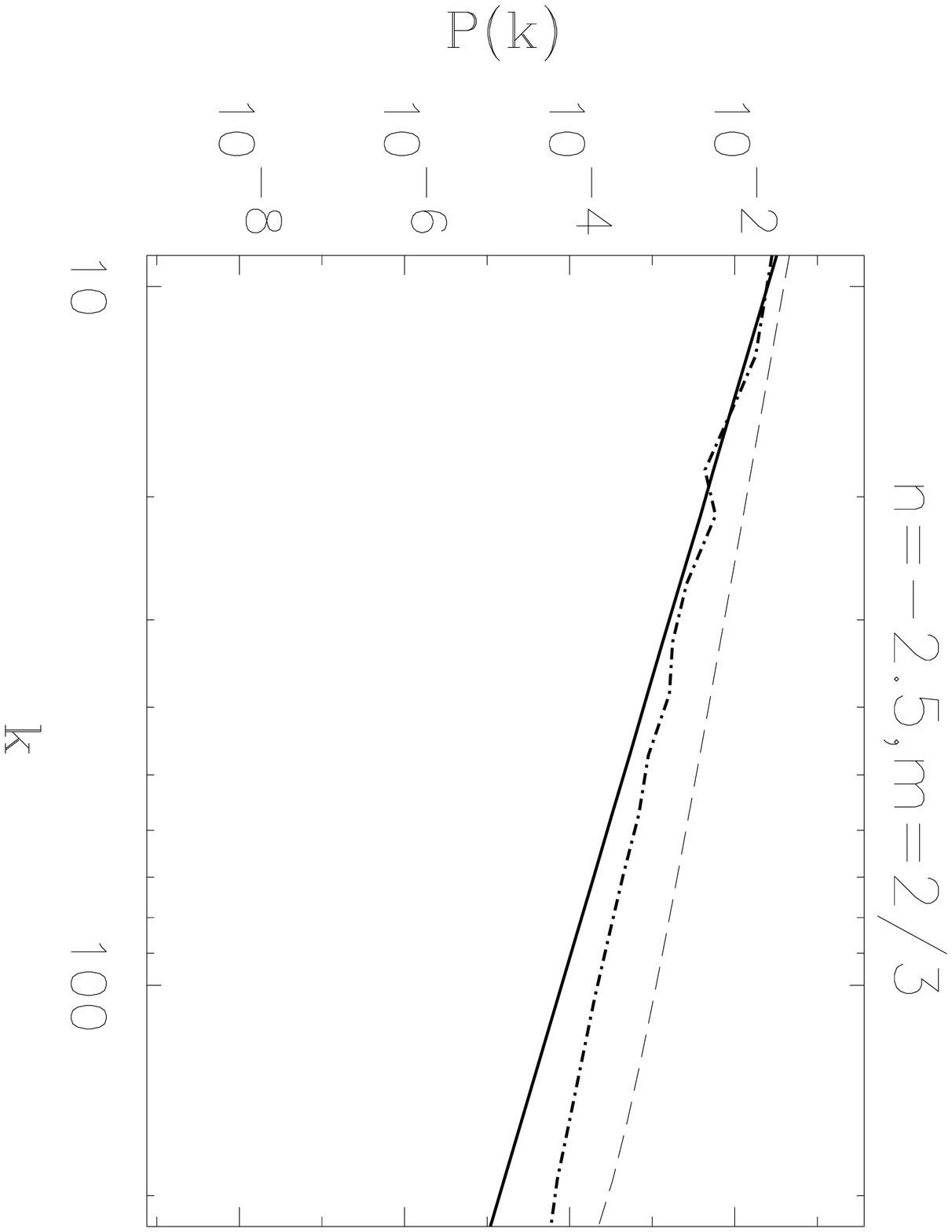} \hfil
\epsfxsize=.25\columnwidth \epsfbox{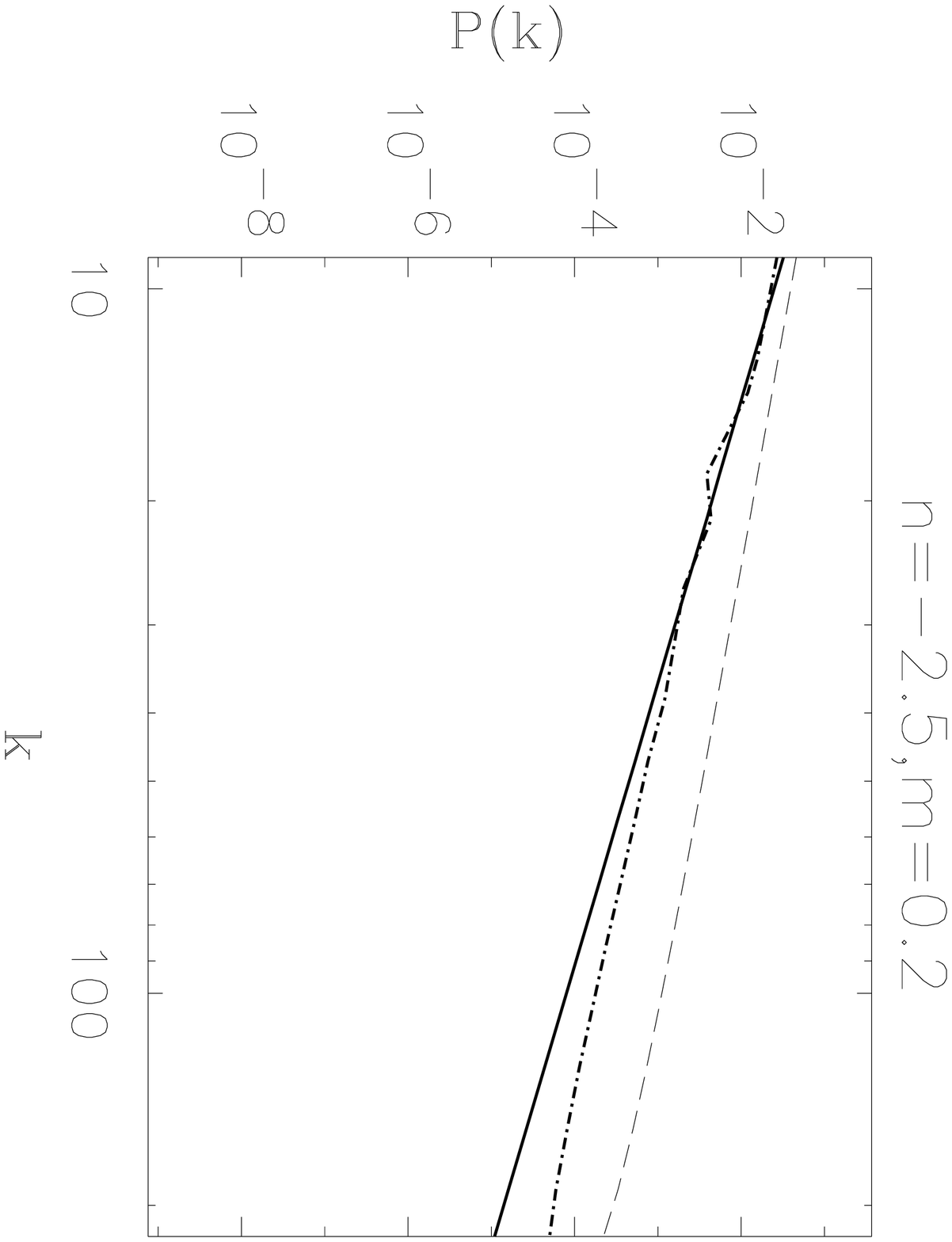}
}
{\centering \leavevmode
\epsfxsize=.25\columnwidth \epsfbox{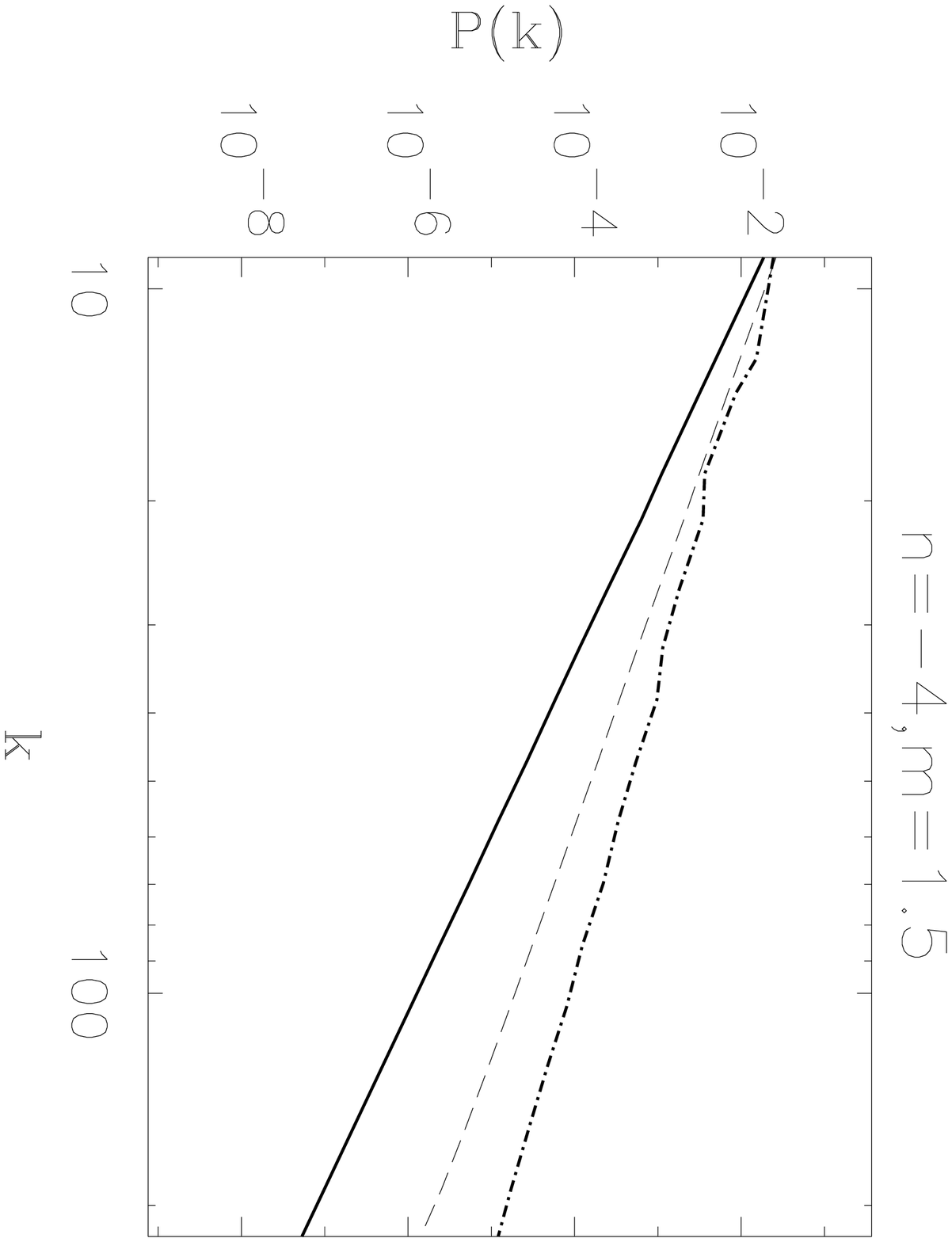} \hfil
\epsfxsize=.25\columnwidth \epsfbox{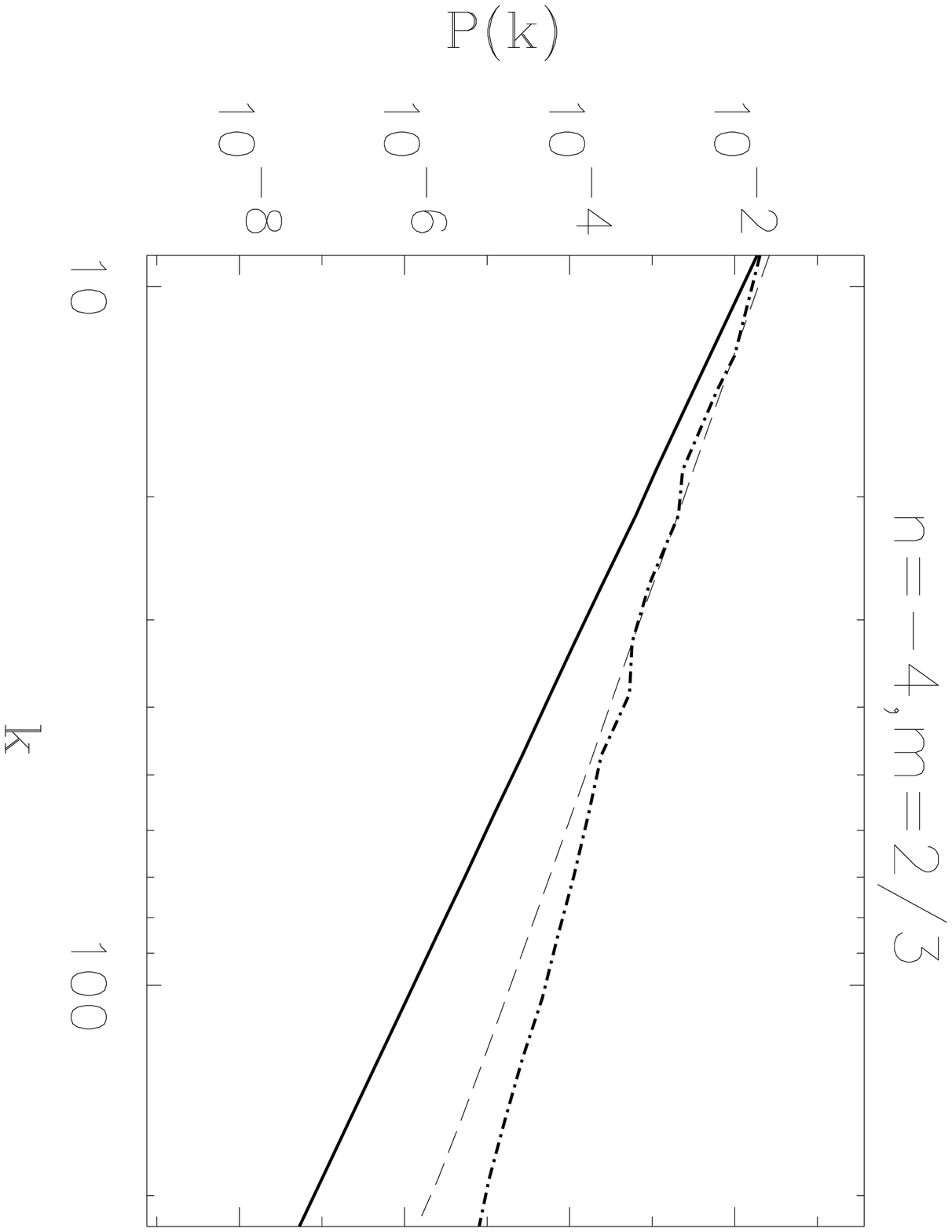} \hfil
\epsfxsize=.25\columnwidth \epsfbox{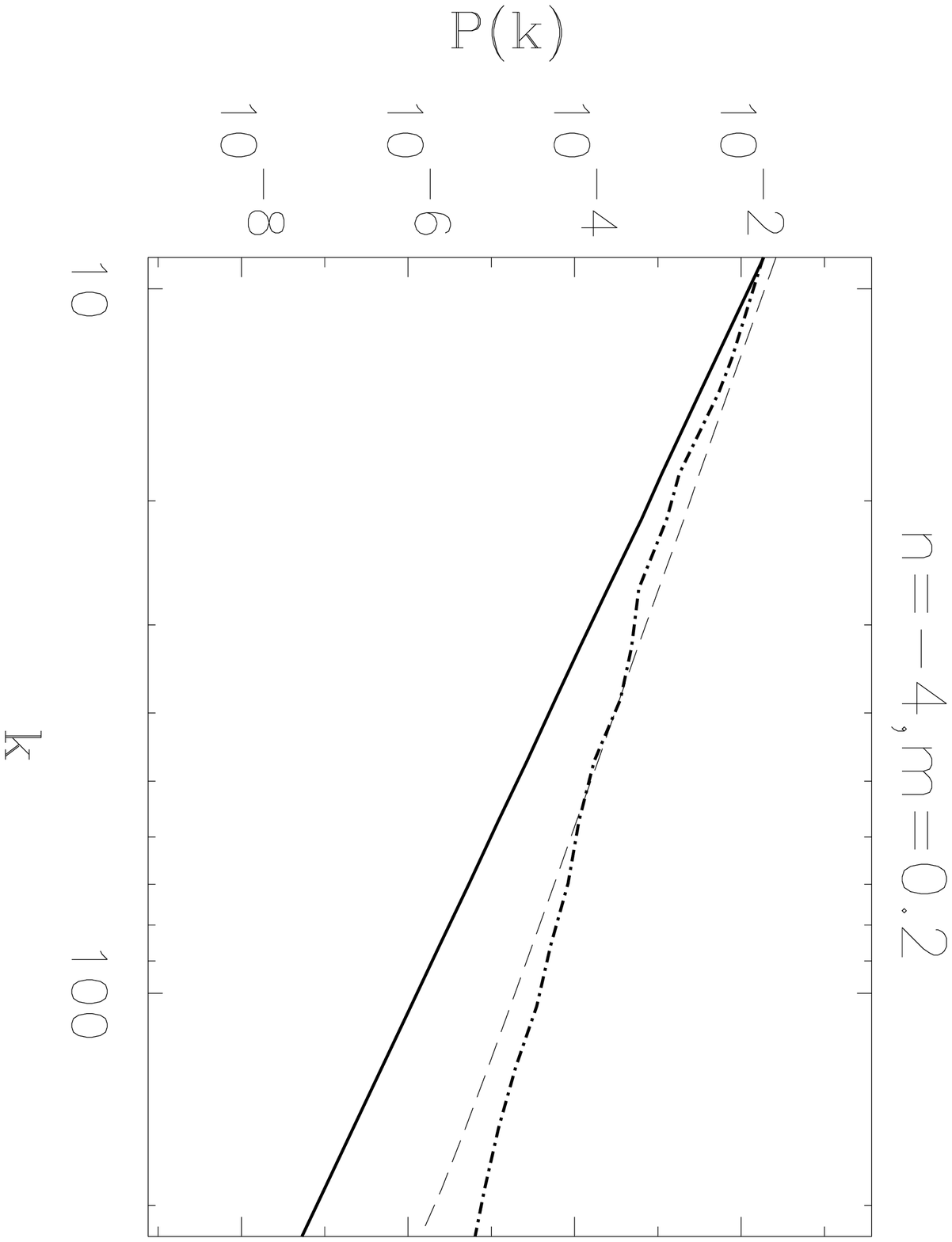}
}
{\centering \leavevmode
\epsfxsize=.25\columnwidth \epsfbox{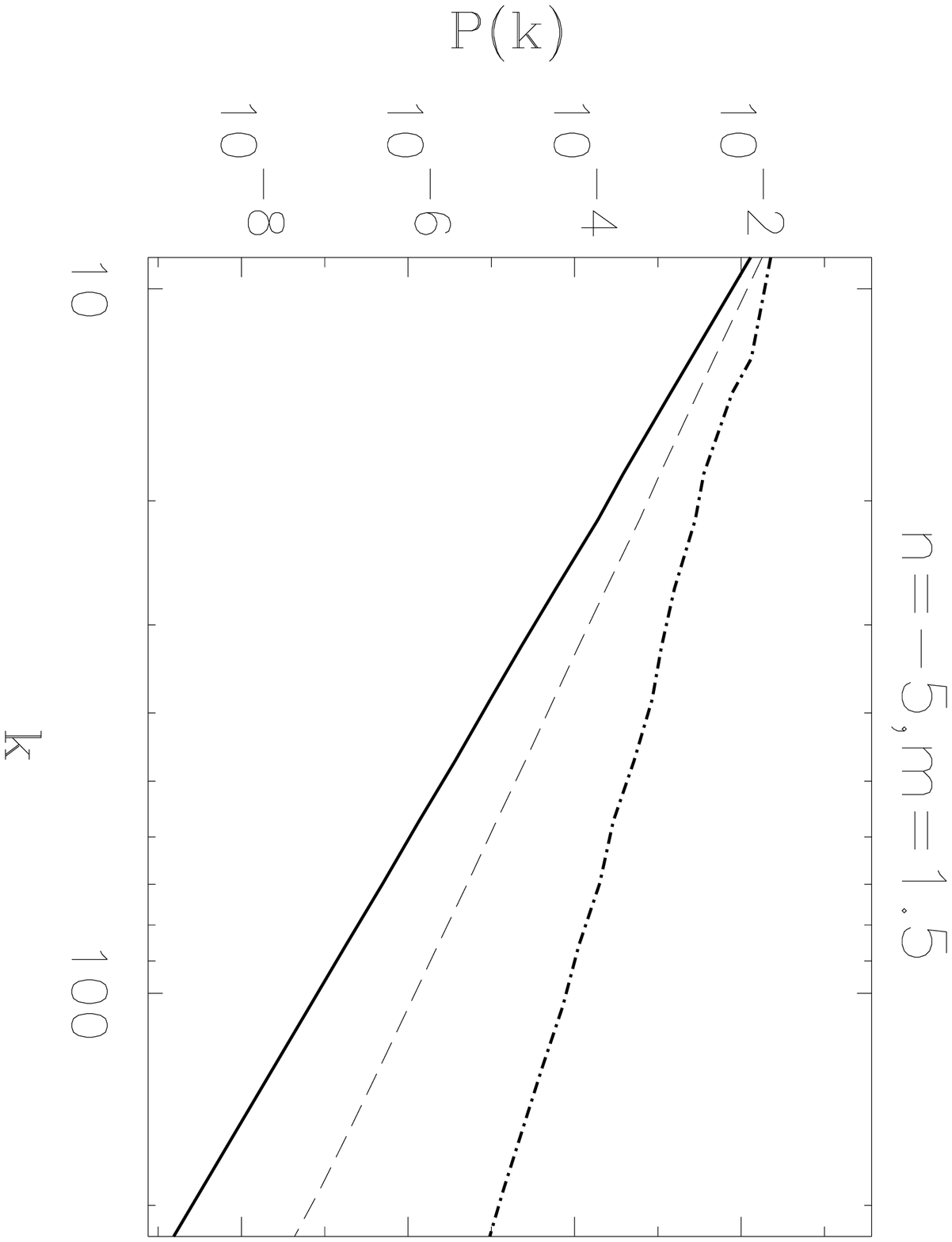} \hfil
\epsfxsize=.25\columnwidth \epsfbox{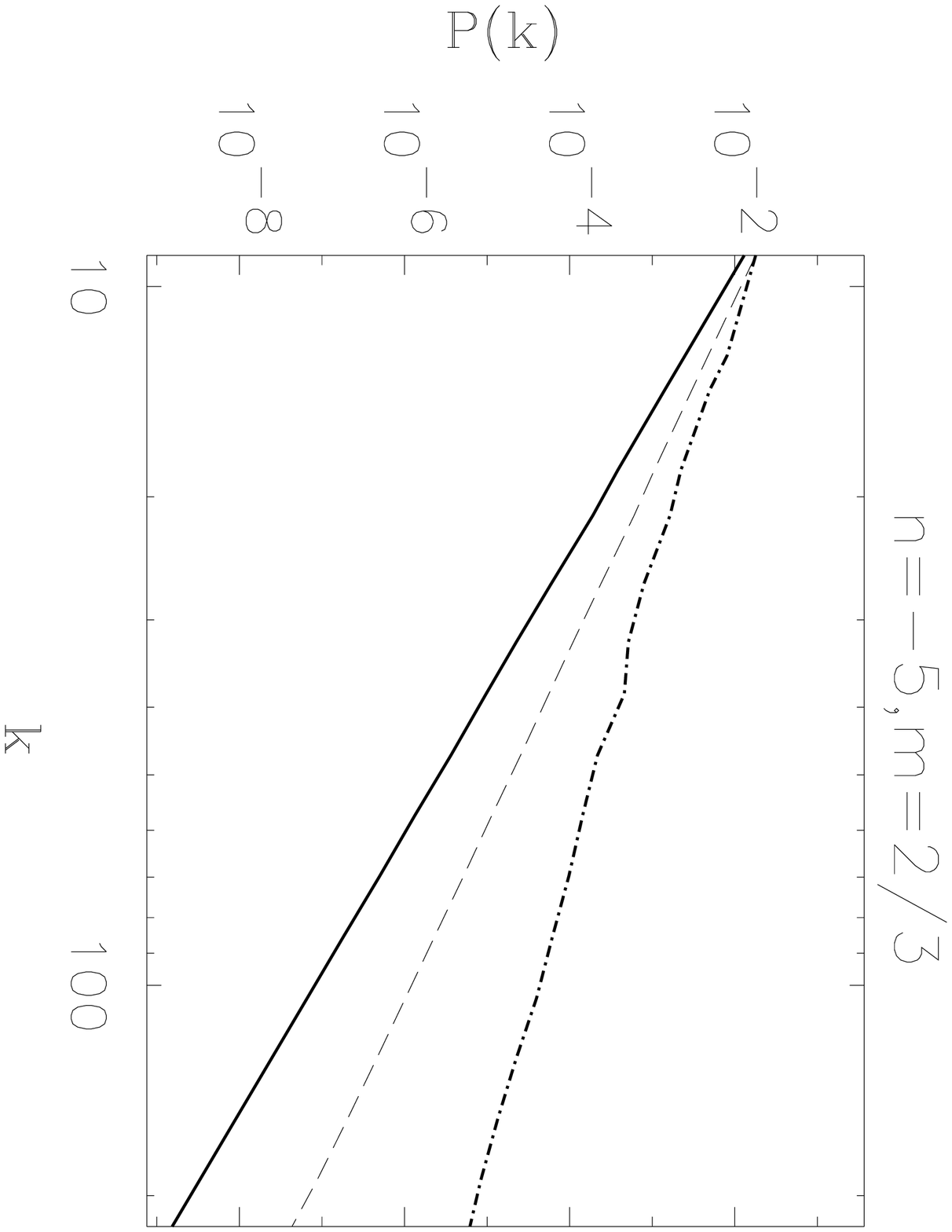} \hfil
\epsfxsize=.25\columnwidth \epsfbox{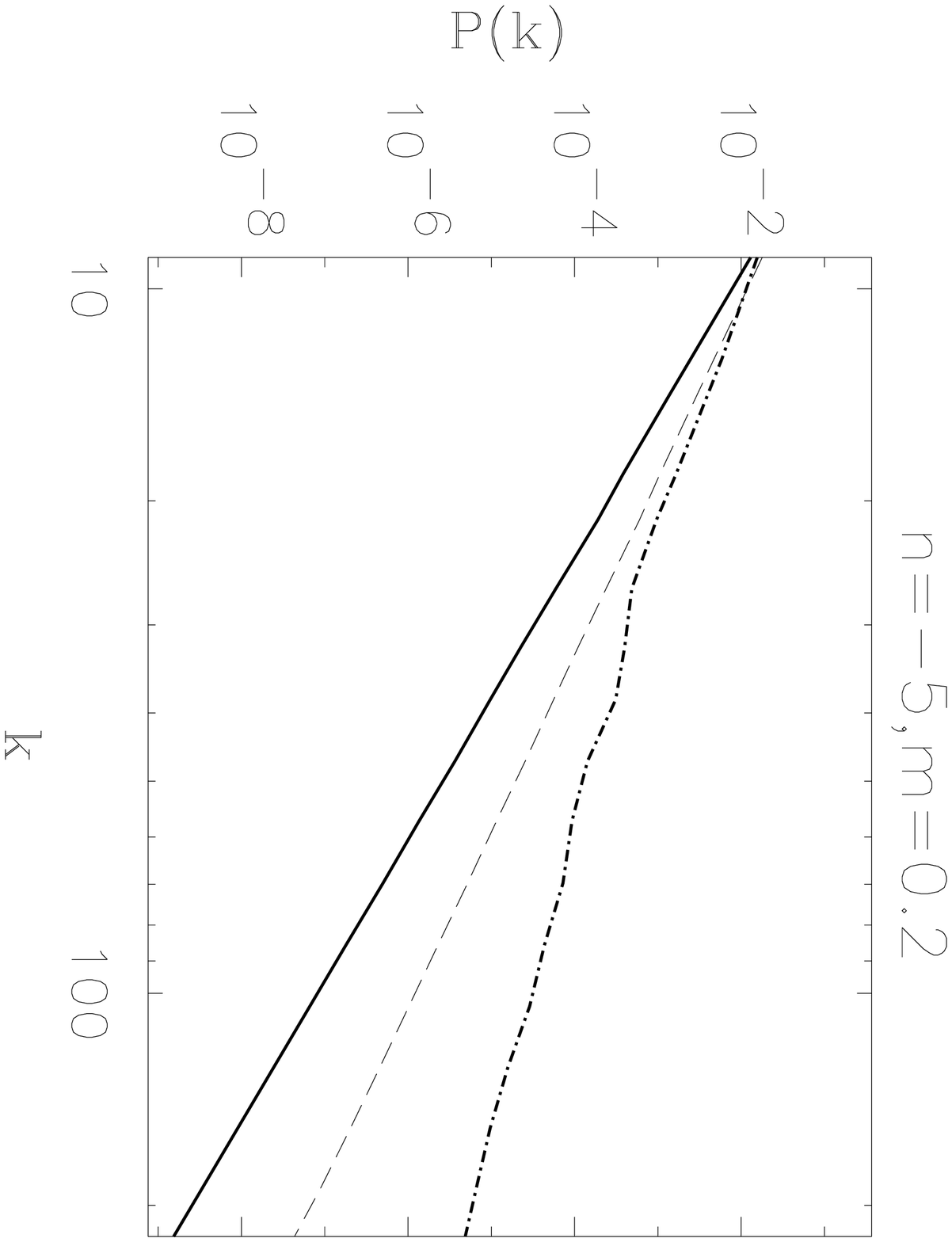} 
}
\caption{
Spectra of the 3D density fields ({\it bold lines}) and of the
emissivity in the 2D channel maps ({\it dash-dotted lines}) for all nine
combinations of density and velocity spectral indices reported in
Table 1. For comparison, the spectra of 2D {\it thin,
spatial} slices of the density are also shown ({\it dashed
lines}). The wavenumber $k$ here is defined as $2 \pi L/\lambda$,
where $L$ and $\lambda$ are as in the caption to fig.\
\ref{fig:old_new_spec}.
}
\label{fig:xi}
\end{figure}

To test the predictions of LP00, in fig.\ \ref{fig:xi} we plot the
spectra of the
emissivity fluctuations in velocity slices of the PPV data cubes (``channel
maps'') ({\it dash-dotted lines}), the spectra of the 3D
density fields ({\it solid lines}), and the spectra of the density in
thin spatial slices of the 3D density field ({\it dashed lines}), for
various combinations of the density and velocity spectral slopes.
These plots confirm the theoretical expectations.
First of all, the transition from thick to thin slices depends
on the spectral index of the velocity.
 For relatively small
numbers of slices we have observed that the slices are still thin
for shallow indices and show signatures of becoming effectively thicker
for steep indices. The width  of the slice is given by
$\Delta V/N$, where $\Delta V=V_{\rm max}-V_{\rm min}$ and $N$ is the
number of slices. The value of $\Delta V$
along the LOS will be a factor of a few higher than the dispersion
$\sigma$ along the same LOS. Conservatively, we shall assume this
factor to be 5, which corresponds to $2.5\sigma$
positive and negative excursions. The squared dispersion
over the box size $L$, $\sigma^2$, equals $CL^m$ for power-law statistics.
Therefore, in order for the slice to be thin, it is required that
$Cr^m>(2.5)^2 CL^m/N^2$ , where $r$ is the scale
under consideration in the plane of the slice, and is essentially the
inverse of wavenumber in the emissivity spectrum.
This means that $N>2.5(L/r)^{m/2}$.
Another expected feature is the flattening of the
spectrum at large $k$ (small separations $r$) due to shot noise.
In our case we derive the slope
considering scales from $r\approx L$ down to $r\approx 0.1 L$, where $L$
is the length of the integration box. Figure
\ref{fig:xi} shows that for $r$ smaller than $0.1 L$ the emissivity
spectrum indeed flattens.

The considerations above  mean
that for the steepest velocity index of $-4.5$ (i.e. $m=1.5$),
the number of slices should be
larger than $30$ in order for them to be thin. On the other hand,
an increase in the number of slices results in higher shot noise due
to the discrete sampling of the velocity along the line of sight (we
only have 100 samples along each line of sight).
Thus, we have adopted a slice width
of $0.03 \Delta V$, which we have found to correspond to
thin slicing for the entire parameter space that we explore, but does
not introduce too high a level of shot noise.

For thin slices and sufficiently {\it steep}
density spectra ($n<-3$), LP00 predicted that the measured emissivity
spectrum {\it does not} depend on the density spectral slope,
and has an index
\begin{equation}
({\it thin~slice~index~if~density~spectrum~is~steep})=-3+m/2.
\label{steep}
\end{equation}
For example, for Kolmogorov
turbulence, $m=2/3$, and therefore an emissivity index of $-2.66$ is
expected. Conversely, for {\it shallow} density spectra, i.e. when
 $n>-3$, the
spectral index in thin slices is expected to be (LP00)
\begin{equation}
({\it thin~slice~index~if~density~spectrum~is~shallow})=n+m/2.
\label{shallow}
\end{equation}
Table~1 provides a quantitative comparison between the theoretical
predictions of LP00 and our numerical results, showing the spectral
indices of the emissivity in the velocity channels as a function of
the indices of the underlying density and velocity fields. The
theoretical predictions of LP00 (given by
equations (\ref{steep}) and (\ref{shallow}))
are indicated in parentheses, while the
measured values in the simulated fields are shown in bold type. It is seen
that in all cases the agreement is within 10\%, for both ``shallow''
and ``steep'' density spectral indices. We stress that the spectra of
the 3D density and 
velocity data cubes, and of the 2D velocity slices, are obtained by
integrating with the appropriate configuration space factors, which are
$r^2dr$ and $rdr$ respectively.

The main prediction of LP00 is the importance of the
velocity statistics in the determination of the emissivity
statistics, specifically, its power spectrum. Our numerical results confirm
this conclusion. Indeed, Table~1 shows that as the velocity spectral
index is varied (downwards along the columns in Table~1), significant
variations occur in the emissivity spectral index. Moreover, for steep
density spectra, the emissivity statistics are roughly independent of
the density spectral index.
For example, consider the cases with velocity indices $-3.2$ and $-11/3$.
It is seen that as the underlying density spectral
index is changed from $-4$ to $-5$, the emissivity spectral index in thin
channels remains constant within the precision of our
measurements. For the case with velocity index of
$-4.5$, the emissivity index changes slightly (from $-2.1$ to $-2.3$)
for the same variation of the density index as above, but such slight
change is within the uncertainty of 0.2, and brackets the value
predicted by the LP00 theory.
{\it By itself, this result sends
a message of
warning against attempts to interpret features in the velocity
channel maps as actual density enhancements.}

\begin{table}[h]
\begin{displaymath}
\begin{array}{lrrr} \hline\hline\\
{\rm Density~index} \rightarrow&  \multicolumn{1}{c}{-2.5~({\rm shallow})}  &
\multicolumn{1}{c}{-4~({\rm steep})} & \multicolumn{1}{c}{-5~({\rm steep})}
\\[1mm]
\hline\\
{\rm Velocity~index}\\
\hfil\downarrow\hfil\\
\hline\hline\\
-3.2 & {\bf -2.2}~(-2.40)& {\bf -2.9}~(-2.90) & {\bf -2.9}~(-2.90)\\[1mm]
\hline \\[1mm]
-11/3 &{\bf -1.9}~(-2.17) & {\bf -2.6}~(-2.67) & {\bf -2.6}~(-2.67)\\[1mm] \hline\\
-4.5 & {\bf -1.8}~(-1.75) & {\bf -2.1}~(-2.25) & {\bf -2.3}(-2.25)\\[1mm] \hline
\end{array}
\end{displaymath}
\caption{Spectral index of the emissivity fluctuations in {\it thin}
velocity channels as a function of the density (top row) and velocity
(left column) fields' spectral indices. The actual measured values are given
in bold font, while the theoretical predictions are given in parentheses.
The differences between the theoretical and observed values are within
the uncertainties of determining the spectral slope. The uncertainty
of the slope fittings is $\sim 0.2$, due to the effects of shot noise
and slight imperfections in the spectrum modification procedure, $\sim
\pm 0.1$ in the imposed spectral index.}
\label{tab:2Dspk_asymp}
\end{table}

The above discussion suggests that, whatever the velocity
spectrum, it can be restored from observations. A possible ambiguity
that arises in the case of shallow density spectra, when
both the velocity and density contribute to the channel map spectrum,
can be resolved by considering several slice widths. For example,
it is obvious that if integration is performed over the whole
velocity range, the resulting statistics can depend on the
density only. The corresponding density inversion is discussed
by Lazarian (1995). For power-law statistics, the spectral index for
{\it very thick} velocity-integrated slices
equals the spectral index of the underlying 3D density.
We have verified this result by integrating over the velocity (the
plots corresponding to the integrated 2D statistics
coincide with those of the 3D spectra in fig.\ \ref{fig:xi}).

We have also confirmed other predictions from LP00, such as the change
of the density spectral index from $n$ in volume to $n+1$ in a thin {\it spatial} density
slice, as shown in fig.\ \ref{fig:xi} by the dashed lines.
  Another important prediction of LP00 was the
gradual steepening of the emissivity spectrum in the velocity slices
as their width increases. This prediction was confirmed by
Stanimirovic (2000) and Stanimirivic \& Lazarian (2000) on the basis
of 21~cm Small Magellanic Cloud data 
and fig.\ \ref{fig:slice_var} shows that
this indeed happens in our simulation data. In particular, the maximum
possible thickening occurs when the velocity information is averaged
out. In our case this corresponds to integration through the velocity
box, or, equivalently, from an integration throughout the full line
of sight in
the original 3D
density data. It is evident that in this case the velocity
fluctuations should not matter and only density statistics should
influence the results. We have verified that both a pure density
integration along the LOS and a velocity integration of the emissivity
data cubes give the same spectral index.

\section{Discussion} \label{sec:disc}

In \S \ref{sec:correl} we have shown that the density and LOS-velocity
fields in our ISM simulation are moderately correlated.
An important question is whether there are interstellar
situations in which the velocity-density correlations are much
stronger than in our simulations and whether the results by LP00 are
applicable to those situations. The answer to this question may
be obtained by analyzing simulations of regions with physical
conditions different from those assumed here.
However, even in the case of
strong correlations, the analytical results of LP00 may hold,
as it was shown by LP00 for a particular case of maximal possible
velocity-density correlations. Further research should clarify the
issue.

Astrophysical implications of the asymptotic relations obtained in
LP00 are discussed at length in Lazarian (1999). Here we will stress two
points. First, the agreement between our measured emissivity spectral
indices and the
predictions of LP00 makes us more optimistic about the application of
the LP00 technique to emission-line studies of actual interstellar
turbulence. Second, it also gives us confidence that the
interpretation by LP00 of the HI intensity spectrum
(Green 1993; Stanimirovic et al.\ 1999) as arising from velocity
fluctuations with a Kolmogorov (i.e., $-11/3$ spectral index) spectrum
is correct.

It is worth pointing out that, even though the notion of
Kolmogorov turbulence has been mentioned frequently throughout the paper, we
do not wish to imply that the ISM in general exhibits Kolmogorov
turbulence. Arguments why
the Kolmogorov (1941) description is not likely to be adequate for the ISM
have been given by several authors (e.g., Scalo 1987; Passot, Pouquet
\& Woodward 1988; Lazarian 1995;
V\'azquez-Semadeni 1999; V\'azquez-Semadeni et al.\ 2000). Briefly
speaking, the ISM flow, unlike
that in the Kolmogorov description, is magnetized and compressible.
One would think that this should change the scaling. Indeed, strong
compressibility may give rise to a
spectrum of shocks, with a slope of $-4$ (in the notation convention
used in this paper). On the other hand,
a theory of magnetized, incompressible turbulence recently
put forward by Goldreich and Sridhar (1995) predicts the same
$k^{-11/3}$ Kolmogorov scaling
for motions perpendicular to magnetic field lines. Such motions would
dominate small-scale asymptotics and therefore the measured spectrum
may stay Kolmogorov-like, although the nature of the cascade is influenced
by magnetic field. Moreover, the interpretation of the observational
data given in LP00, and supported in this paper,
is suggestive that the Kolmogorov scaling may ultimately survive
compressibility or, alternatively, that the HI data considered there
sample gas that is only weakly compressible. Further work, both theoretical and
observational, is needed to resolve this issue. In particular,
application of the techniques used here to molecular-line data should prove
of great interest.

To conclude we should stress that for Kolmogorov turbulence there
exists a {\it coincidence} between the spectral index of a thin,
2D spatial slice of the
density, which is:
\begin{equation}
({\it spectral~index~of~thin~2D~density~slice})=n+1
\label{2Ddensity}
\end{equation}
and the spectral index  of the emissivity fluctuations in a thin
velocity channel when the density spectrum is steep
(e.g. Kolmogorov) and the velocity spectrum follows a Kolmogorov law.
Indeed, the former,
calculated using eq.~(\ref{2Ddensity}), is $-11/3+1=-8/3$
while the latter, calculated using
 eq.~(\ref{steep}), takes the same value $-3+1/3=-8/3$. Although
the accidental character of this coincidence was mentioned in LP00,
our experience shows that it does cause confusion.
For instance, we have had to confront the point of view
that the observed spectrum of HI fluctuations is a simple consequence
of eq.~(\ref{2Ddensity}). This is a {\it fallacy} and, while we do not wish to
repeat the arguments of LP00, Table~1
shows that the naive ``rule'' given by eq.~(\ref{2Ddensity})
does not return the correct emissivity spectral indices
for velocity spectra other than Kolmogorov's, and that the coincidence is
accidental for the Kolmogorov index. The correct
formula for calculating the spectral index of the emissivity fluctuations in
thin slices when the density spectrum is steep
is given by eq.~(\ref{steep}).

\section{Summary and implications} \label{sec:sum}

In this paper we have used data from a numerical simulation of
compressible MHD turbulence in the ISM to study the correlation
between the density and LOS-velocity field, and the dependence
of the spectral index of the velocity-channel emissivity on the indices of
the original three-dimensional velocity and density fields. To do so,
we have {\it a)} computed the cross-correlation between the density and
LOS-velocity fields, {\it b)} modified their spectral slopes to
pre-determined values, and {\it c)} produced PPV data from them.

 From the PPV data, we computed in turn the emissivity spectrum in velocity
slices of those cubes (``channel maps''), allowing us to directly test the
results from the analytical study by LP00. We have found
that its predictions hold also for the case of fields obeying
the fluid conservation equations.
In particular, we have found that for steep
density spectra with power-law indices $n<-3$, the emissivity spectral index
does not depend on the actual value of the density index and is
determined exclusively by the velocity fluctuations. This indicates
that only the velocity field is
responsible for the structure existing in thin slices of the
PPV cubes if the density spectrum is steep enough, in agreement with the
previous result by Pichardo et al.\ (2000) that the morphology in the
channel maps is more resemblant of that of the LOS-velocity field than
that of the density field, and with previous suggestions that
``objects'' identified in position-velocity maps may not correspond to
actual density features (Adler \& Roberts 1992; Pichardo et al.\ 2000;
Ostriker, Stone \& Gammie 2000).

\begin{figure}[ht]
{\centering \leavevmode
\epsfxsize=.45\columnwidth \epsfysize=.45\columnwidth
\epsfbox{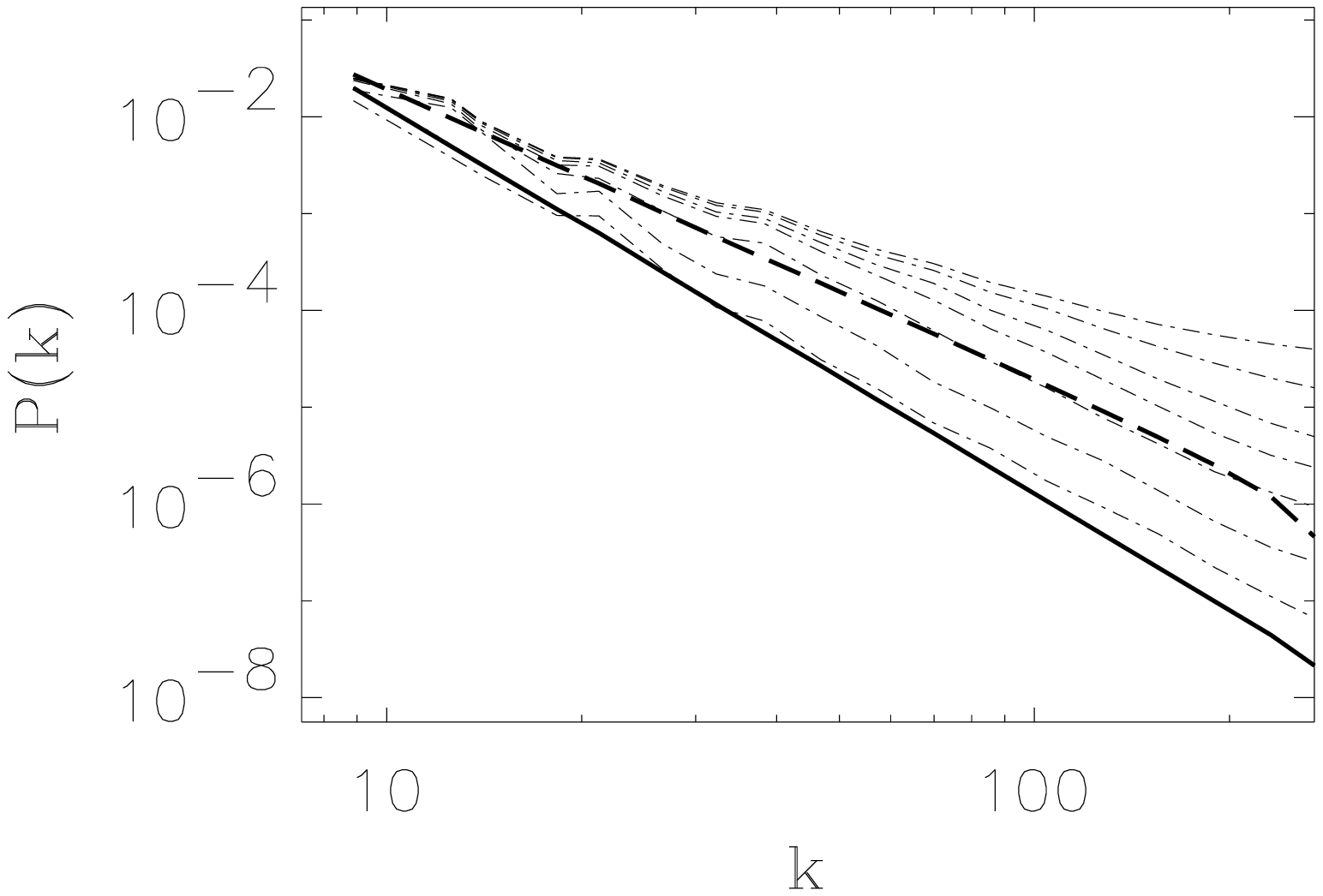} \hfil
\epsfxsize=.45\columnwidth \epsfbox{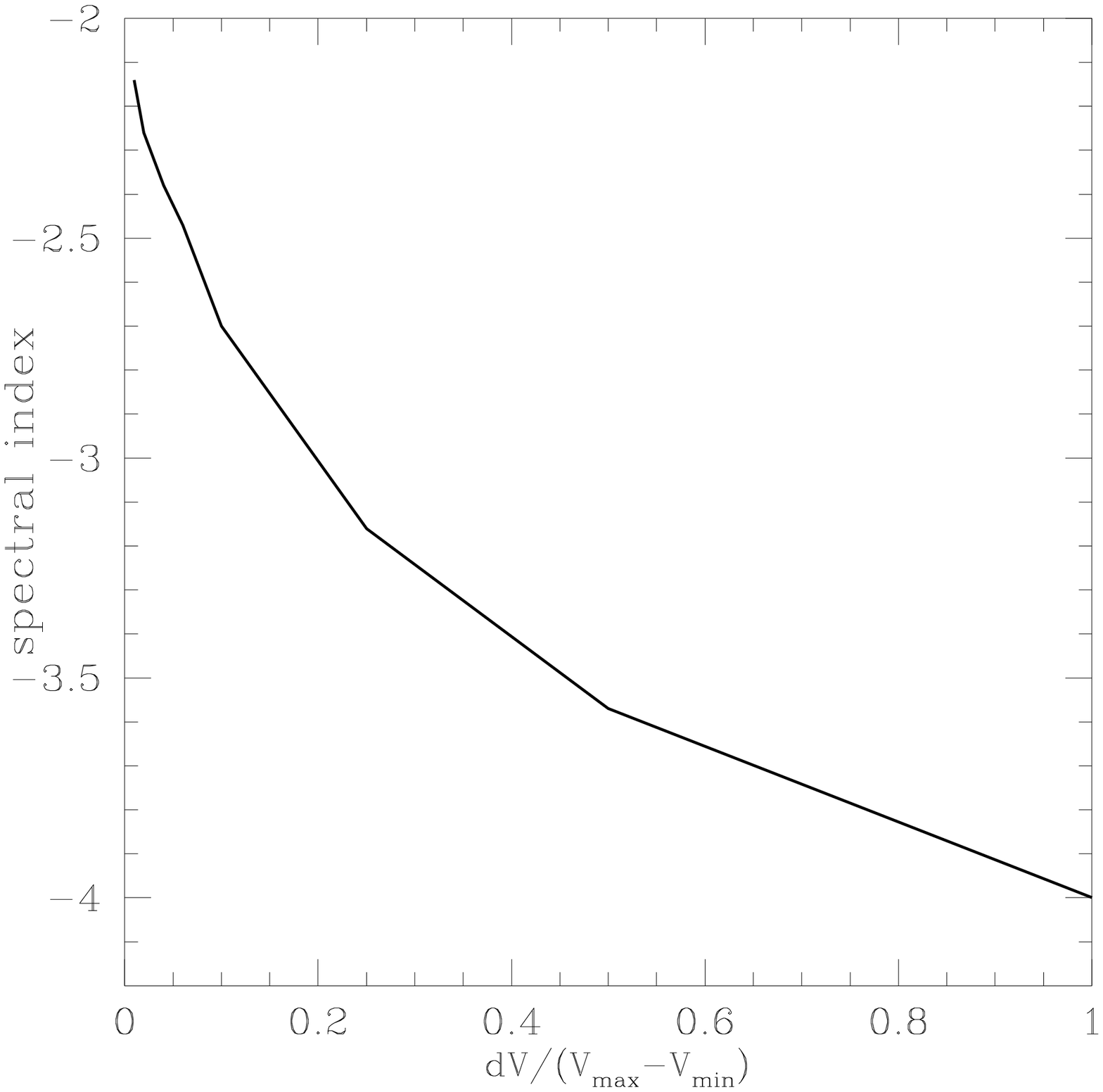} \hfil
}
\caption{{\it Left Panel:} {\it Thin dash-dotted lines:} emissivity
spectra for velocity slices of 
various widths, for velocity spectral index $-4.5$ and density
index $-4$. From top to bottom, the curves correspond to
slices of width $0.01$, $0.02$, $0.04$,
$0.06$, $0.1$ $0.25$, and $0.5$ times $\Delta V$. {\it Thick solid 
line:} power spectrum of the LOS-projected density field (i.e., column
density). The emissivity spectra in the velocity slices are seen to
approach the spectrum of column density as they get wider. {\it Thick
dashed line:} power spectrum of a {\it thin} (1 grid cell) spatial density
slice. The emissivity spectra in thin velocity slices are seen to be
strongly influenced by the velocity field, as they have different
slopes than the spectra of either the column density or the thin
density spatial slice. {\it Right Panel:} Variation
of the emissivity spectral index as a function of the velocity slice
width for the same data cube.}
\label{fig:slice_var}
\end{figure}

The spectral index measured for the emissivity fluctuations in the
numerical simulation
was within 10\% of the value given by relation (\ref{steep}),
derived by LP00, in which $m$ is
the exponent of the second-order structure
function. For a shallow density spectrum, namely
$n>-3$, we found that the emissivity index depends on both velocity and
density. Again the LP00 expression for the resulting power-law index,
relation (\ref{shallow}), provides a satisfactory description of
our measurements.
We also estimated the accuracy of our measurements both by integrating out the
velocity within PPV data cubes and by taking thin slices of the density
data cubes. These cases have straightforward analytical solutions (see
LP00) and served as benchmarks.
We also verified that variations of the emissivity spectrum
resulting from the change of the velocity slice width (see fig.\
\ref{fig:slice_var}) are very
similar to those reported in the studies by Stanimirovic (2000) and
Stanimirovic \& Lazarian (2000),
in which the width of observational HI data slices was varied to
test the predictions from LP00.


In brief, our results imply that:\\
1. Velocity creates small scale structure within slices of PPV data cubes
and therefore the interpretation of features seen in PPV data cubes
as density structures (``clouds'') may be misleading.\\
2. For sufficiently shallow density spectra, velocity dominates the
statistics of intensity fluctuations within PPV slices.\\
3. Emissivity spectra become steeper as the width of the velocity
slices increases
and, for sufficiently thick slices, the spectra are dominated by
density enhancements only.

\acknowledgements

Part of this work was completed while two of us (A.L.\ and E.V.-S.)
participated in the Astrophysical Turbulence Program of the Institute
for Theoretical Physics at the University of California at Santa
Barbara. The numerical simulation was performed on the Cray Y-MP 4/64
of DGSCA, UNAM. This work has received partial funding from CONACYT, M\'exico,
through grant 27752-E to E.V.-S.\ and from NSF, USA, through grant PHY94-07194.

\clearpage

\end{document}